\documentclass[aps,prx,twocolumn]{revtex4-1}
\usepackage{amscd,amssymb,amsmath,latexsym,bm}
\usepackage[mathcal,mathscr]{euscript}
\usepackage{lipsum}
\usepackage{amsfonts}
\usepackage{amsmath}
\usepackage{amssymb}
\usepackage{txfonts}
\usepackage{pxfonts}
\usepackage{graphicx}
\usepackage{bm,units,yfonts}
\usepackage[table,dvipsnames]{xcolor}
\usepackage{hyperref}
\usepackage{lineno}

\newcommand{\CM}{{\mathbb C}}

\newcommand{\PM}{{\mathbb P}}
\newcommand{\RM}{{\mathbb R}}
\newcommand{\SM}{{\mathbb S}}
\newcommand{\TM}{{\mathbb T}}
\newcommand{\ZM}{{\mathbb Z}}

\newcommand{\Aa}{{\mathcal A}}

\newcommand{\Hh}{{\mathcal H}}

\newcommand{\Nn}{{\mathcal N}}

\newcommand{\Ss}{{\mathcal S}}

\newcommand{\Ll}{{\mathcal L}}
\definecolor{bleudefrance}{rgb}{0.19, 0.55, 0.91}

\begin{document}
	
	\title{Mapping Chern Numbers in Quasi-Periodic Interacting Spin Chains}
	
	\author{Yifei Liu, Emil Prodan}
	
	\affiliation{
		Department of Physics, Yeshiva University, New York, NY 10016, USA
	}
	
	\begin{abstract} Quasi-periodic quantum spin chains were recently found to support many topological phases in the finite magnetization sectors. They can simulate strong topological phases from class A in arbitrary dimension, characterized by first and higher order Chern numbers. In the present work, we use those findings to generate topological phases at finite magnetization densities that carry first Chern numbers. Given the reduced dimensionality of the spin chains, this provides a unique opportunity to investigate the bulk-boundary correspondence as well as the stability and quantization of the Chern number in the presence of interactions. The later is reformulated using a torus action on the algebra of observables and its quantization and stability is confirmed by numerical simulations. The relations between Chern values and the observed edge spectrum are also discussed.
	\end{abstract}
	
	\maketitle
	
	
	
	
	\section{Introduction}
	
	Topological gaps that host robust edge modes can be easily generated using aperiodic patterning. This has been now demonstrated with condensed matter systems~\cite{YoshidaPRB2013,HeEPL2015,ProdanPRB2015,
		TranPRB2015,VidalPRB2016,FulgaPRL2016,
		CollinsNature2017,AgarwalaPRL2017,HuangPRL2018,BournJPA2018,
		VarjasPRL2019,DevakulPRB2019,PaiPRB2019,KellendonkAHP2019,
		ChenArxiv2019,IliasovPRB2020,FremlingPRB2020,
		HuangPRB2020,ChenPRL2020,DuncanPRB2020}, photonic systems~\cite{KrausPRL2012,VerbinPRL2013,Vardeny2013,
		TanesePRL2014,VerbinPRB2015,HuPRX2015,BandresPRX2016,
		DareauPRL2017,BabouxPRB2017,ZilberbergNature2018,
		KollarNature2019,KollarCMP2020,CarusottoNatPhys2020,
		SchultheissAPX2020, YangLSA2020,ZhouLSA2020}, acoustic systems~\cite{ApigoPRL2019,NiCP2019,ChengPre2020} and mechanical systems~\cite{MitchellNature2018,MartinezPTRSA2018,ApigoPRM2018,RosaPRL2019,PalNJP2019, ZhouPRX2019,XiaPRAppl2020,RivaPRB2020,RivaPRB2020b,
		XiaArxiv2020,RosaArxiv2020}. The works we just mentioned dealt only with the un-correlated states of condensed matter systems or with the linear regime of the classical systems. In these settings, most of the difficult questions such as the robustness of the topological invariants \cite{BellissardJMP1994,ProdanPRL2010,LoringEPL2011,ProdanJPA2013,
		ShemPRL2014,ThiangAHP2016,ProdanJFA2016,BourneRMP2016,BourneMPAG2018,BourneJPA2018} or the bulk-boundary principles \cite{KRS-BRevMathPhys2002,ProdanSpringer2016,BourneAHP2017,BourneAHP2019} in arbitrary dimensions and for arbitrary patterns of atomic configurations are now well understood and experimentally under control. A vigorous research is currently underway on the interplay between the aperiodicity, many-body correlations and topology in quantum systems or between aperiodicity, non-linear effects and topology in classical systems \cite{HePRA2013,HuPRB2016,ZengPRB2016,LiEPL2017,KunoNJP2017,
		MarraEPJ2017,TaddiaPRL2017,KePRA2017,NakagawaPRB2018,
		SarkarSR2018,HuPRA2019,LadoPRR2019,OritoPRB2019,
		ZuoNJP2020,ChenPRA2020,RosnerArxiv2020}. At the mathematically rigorous level, there have been exciting new developments \cite{Hastings2015,GiulianiPRB2016,BachmannAHP2018,GiulianiJSP2020,
		LuAOP2020,BachmannCMP2020} on the formal definition and quantization of the linear transport coefficients. However, there are no results for the non-linear transport coefficients that connect to higher Chern numbers and the bulk-boundary remains poorly understood in the correlated case. Furthermore, people are interested in a much more ambitious program whose goal is identifying all topological invariants that can be associated to a given class of many-body models. The progress for this program can be accelerated if more manageable models become available, especially if these models can be prescribed with non-trivial topology and if they can be simulated on a computer.   
			
The purpose of our present work is to supply a high throughput of interacting models where the existing predictions can be tested and the bulk-boundary correspondence can be studied. For example, the reader will see Theorem~1 from \cite{LuAOP2020} in action for eight distinct topological many-body gaps and varying pseudo-fluxes. The many-body models derived here are based on our recent work \cite{LiuArxiv2020}, where we detected a large family of topological phases in quasi-periodic XXZ-spin models by using K-theoretic methods \cite{Bellissard1986,Bellissard1995,KellendonkRMP95,ProdanJGP2019}. The analysis in \cite{LiuArxiv2020} was restricted to invariant sectors with finite but otherwise arbitrary total magnetization $M$. In the present work, we use the topological states discovered in \cite{LiuArxiv2020} inside the $M=1$ sector, to generate new gapped many-body states that carry non-trivial Chern numbers, this time at finite magnetization densities. We also succeed in generating gapped many-body states from the higher magnetization sectors but those results will be reported somewhere else.  

Another objective of the paper is to report an alternative way of defining and computing a Chern number for spin systems. Let us recall that a standard approach in the simulation of the topological phases \cite{ShengPRL2003,WanPRB2005,KudoPRL2019} is to examine the vector bundle traced by the ground state under twisted boundary conditions. This method, for example, was recently used in \cite{HuPRB2016} on a quasi-periodic spin chain to detect a fractional Chern state. While a very convenient computational tool, the approach has two main shortcomings: 1) By its very nature, the technique requires a boundary, hence the topological invariants cannot be defined directly in the thermodynamic limit, as it is the case for the rigorously defined ones \cite{BellissardJMP1994} in the uncorrelated case. As such, it is impossible to prove the convergence or to find the rate of convergence of the finite size calculations, as it was done in the uncorrelated case \cite{ProdanSpringer2017}. 2) When the sites carry internal degrees of freedom, such as spin, the twisted boundary conditions are not uniquely defined. For example, the spin-Chern number \cite{ShengPRL2006,ProdanPRB2009} can be defined by a proper choice of the twisted boundary condition.

A powerful and extremely general framework for generating topological invariants on algebras of quantum observables is the local index formula for semi-finite spectral triples \cite{CareyAMS2014}. As explained in Sec.~5.1 of \cite{CareyAMS2014}, such spectral triples can be generated from torus actions and invariant trace states on the algebra of observables. A physically transparent torus action on the algebra of spin operators is described in our Section~\ref{Sec:Chern}. We use it to reformulate the Chern number such that the expression holds in the thermodynamic limit and fits into the framework developed in \cite{CareyAMS2014}. A more mathematical analysis will be presented elsewhere and here we focus on the numerical evidence supporting our proposed definition. For this, we evaluate the proposed Chern numbers for un-correlated and correlated ground states and show that they agree with the expected quantized values. We also investigate the edge spectra of the models, which are all un-gapped, as expected.

The paper is organized as follows. Sec.~\ref{Sec:AperiodicSpinChains} introduces the aperiodic spin model used in all our numerical simulations and another, more general, class of aperiodic models used in the theoretical arguments. The fermion version of the models is also discussed via the Jordan-Wigner mapping. The latter is only used as guidance in our theoretical arguments. Sec.~\ref{Sec:M1Jz0} supplies an analysis of the magnetization sector $M_z=1$, where ordinary topological gaps are detected and characterized. Sec.~\ref{Sec:MBodyGaps} uses the results in the previous section to construct gapped many-body states at finite magentization densities. Sec.~\ref{Sec:Chern} presents our new expression for the Chern number and reports numerical evaluations of it for the gapped states engineered in the previous section. The paper ends with a section focused on the interpretation of the results and outlooks.
	
	\section{Quasi-periodic Spin Chains}
	\label{Sec:AperiodicSpinChains}
	
In this section, we describe a simple family of quasi-periodic patterns and introduce the quantum spin models defined over these patterns that will be used in our numerical simulations. To support our theoretical arguments, we also introduce a more general family of spin Hamilotonians, for which we demonstrate a covariant property and discuss its implications.
	
	\subsection{The spin system defined}
	
	We consider a spin-$\frac{1}{2}$ chain over a 1-dimensional lattice of points $\Ll = \{p_n\}_{n =\overline{-L,L}}$ generated with the algorithm
	\begin{equation}\label{Eq:PattAlg}
	p_n = n + 0.45\, \big (\sin[2 \pi (n\theta + \varphi)] -\sin(2 \pi \varphi)\big ),
	\end{equation} 
	where the parameters belong to the circle, $\theta$, $\varphi \in \SM \ (= \RM/\ZM)$. Eq.~\ref{Eq:PattAlg} was written such that $p_0$ sits at the origin. Throughout, we will use the notation $|\cdot|$ to indicate the cardinal of set. As such, the size of the pattern is $|\Ll|=2L+1$. At many places, this size will be considered infinite.
		
Our main focus is on the cases when $\theta$ is fixed at irrational values and the lattice is truly aperiodic. For a spin system over such pattern, it is natural to assume that the spin-spin interaction depends on the separation distance $d_n = p_{n+1}-p_n$ between the points. As such, we consider spin Hamiltonians of the form
\begin{align}\label{Eq:GenHam1}
	H_S = \sum_{n = -L}^L & \left [J(d_n) \big (S^{x}_n S^x_{n+1} + S^{y}_n S^y_{n+1} \big ) \right . \\ \nonumber
	 & \quad \left . + J_z (S^{z}_n + \tfrac{1}{2}) (S^z_{n+1}+\tfrac{1}{2}) \right ] .
\end{align}
The operator corresponding to the $z$-component of the magnetization
	\begin{equation}
	M_z = \sum_{n=-L}^L \big ( S_n^z+\tfrac{1}{2} \big )
	\end{equation}
commutes with the Hamiltonians \eqref{Eq:GenHam1}. The eigenvalues of $M_z$ will be denoted by $M$ and the corresponding eigen-spaces will be referred to as $M$ sectors.

In all our numerical simulations, we use the models \eqref{Eq:GenHam1} with finite chain size, and both closed and open boundary conditions are used. To accommodate the close boundary condition, we only sample the quantized values $\theta_k = k/|\Ll|$, $k=1,\ldots,|\Ll|$. All numerical results reported here are obtained with the choice 
	\begin{equation}
	J(s) = e^{-|s|}, \quad s \in \RM,
	\end{equation}
but, as we shall see below, our main conclusions are independent of the concrete functional form. As for the Ising interaction strengths, we set them from the beginning to be independent of $d_n$. 

When the arguments assume an infinite spin chain, the Hamiltonians $H_S$ can be viewed as derivations on the algebra of local spin observables, as explained in \cite{BratelliBook2}[Sec.~6.2]. They can also be viewed as unbounded operators on the Hilbert space generated by the GNS representation corresponding to the state with all the spins down (see \cite{BratelliBook2}[p.~349]). Explicitly, this Hilbert space is $\Hh= \bigoplus_{M=0}^\infty \Hh_M$, where $\Hh_M$'s represent the magnetization sectors. We will always specify which point of view is adopted in our arguments.
	
\subsection{Connections with fermionic models}
	
	We will use the Jordan-Wigner mapping \cite[Sec.~5.1]{ColemanBook} to make connections with fermionic aperiodic physical systems, which will supply guidance at several places. The mapping is defined by the operators
	\begin{equation}\label{Eq:StoA}
	a_n = S_n^- \prod_{j=-L}^{n-1} 2S_j^z, \quad a_n^\ast = S_n^+ \prod_{j=-L}^{n-1} 2S_n^z,
	\end{equation}
	which satisfy the canonical anti-commutation relations
	\begin{equation}\label{Eq:Rel2}
	a_m a_n + a_n a_m=0, \quad a_m^\ast a_n + a_n a_m^\ast = \delta_{m,n}.
	\end{equation} 
	Together with the inverse formulas
	\begin{equation}\label{Eq:AtoS}
	S_n^{-} = a_n \prod_{j=-L}^{n-1}(2 a_j^\ast a_j -1), \ \  S_n^{+} = a_n^\ast \prod_{j=-L}^{n-1}(2 a_j^\ast a_j -1),
	\end{equation}
	these relations establish an isomorphism between the algebra of spin operators and the algebra of fermionic creation and annihilation operators over the finite lattice $\Ll$. In particular, the magnetization operator is mapped into the operator of particle number
	\begin{equation}\label{Eq:MtoN}
	M_z \rightarrow N=\sum_{n=-L}^L a_n^\ast a_n,
	\end{equation}
	and the model Hamiltonian \eqref{Eq:GenHam1} into \cite[p.~74]{ColemanBook}
	\begin{align}\label{Eq:FHam}
	H_F = \sum_{n = -L}^L & \big [\tfrac{1}{2} J(d_n) (a^\ast_n a_{n+1} + a^\ast_{n+1} a_{n} )  +J_z a_n^\ast a_n a_{n+1}^\ast a_{n+1} \big ].
\end{align}
This fermionic Hamiltonian will only be used in theoretical arguments. All numerical simulations reported below were performed with the spin Hamiltonian~\eqref{Eq:GenHam1}.

\subsection{A Generic Family of Spin Hamiltonians}
\label{Subsec:Algebra}

\begin{figure}[t!]
		\includegraphics[width=\linewidth]{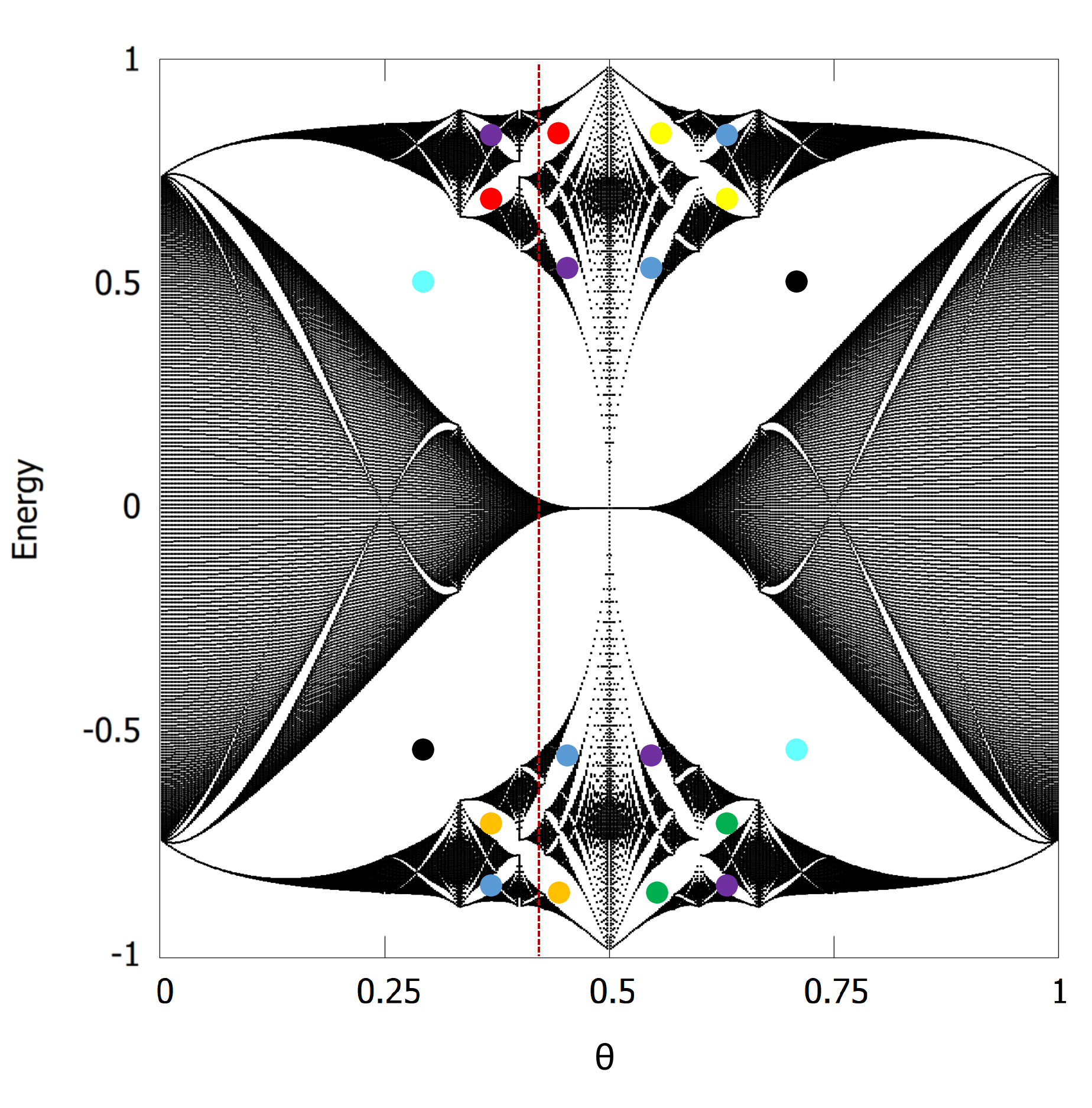}
		\caption{\small Energy spectrum of Hamiltonian~\eqref{Eq:GenHam1} inside the $M=1$ sector as function of $\theta$. The simulations were performed with $J_z=0$, $|\Ll|=501$ and the range of parameter $\theta$ has been sampled at rational values $\theta_k = \frac{k}{|\Ll|}$ to accommodate closed boundary conditions. Eight gaps are identified and color coded for future references. The vertical dotted line indicates the $\theta$ where the bulk-boundary principle is probed in Fig.~\ref{Fig:M1JZ0Phi}.}
		\label{Fig:M1JZ0Spec}
\end{figure}
	
\begin{figure}[t!]
		\includegraphics[width=\linewidth]{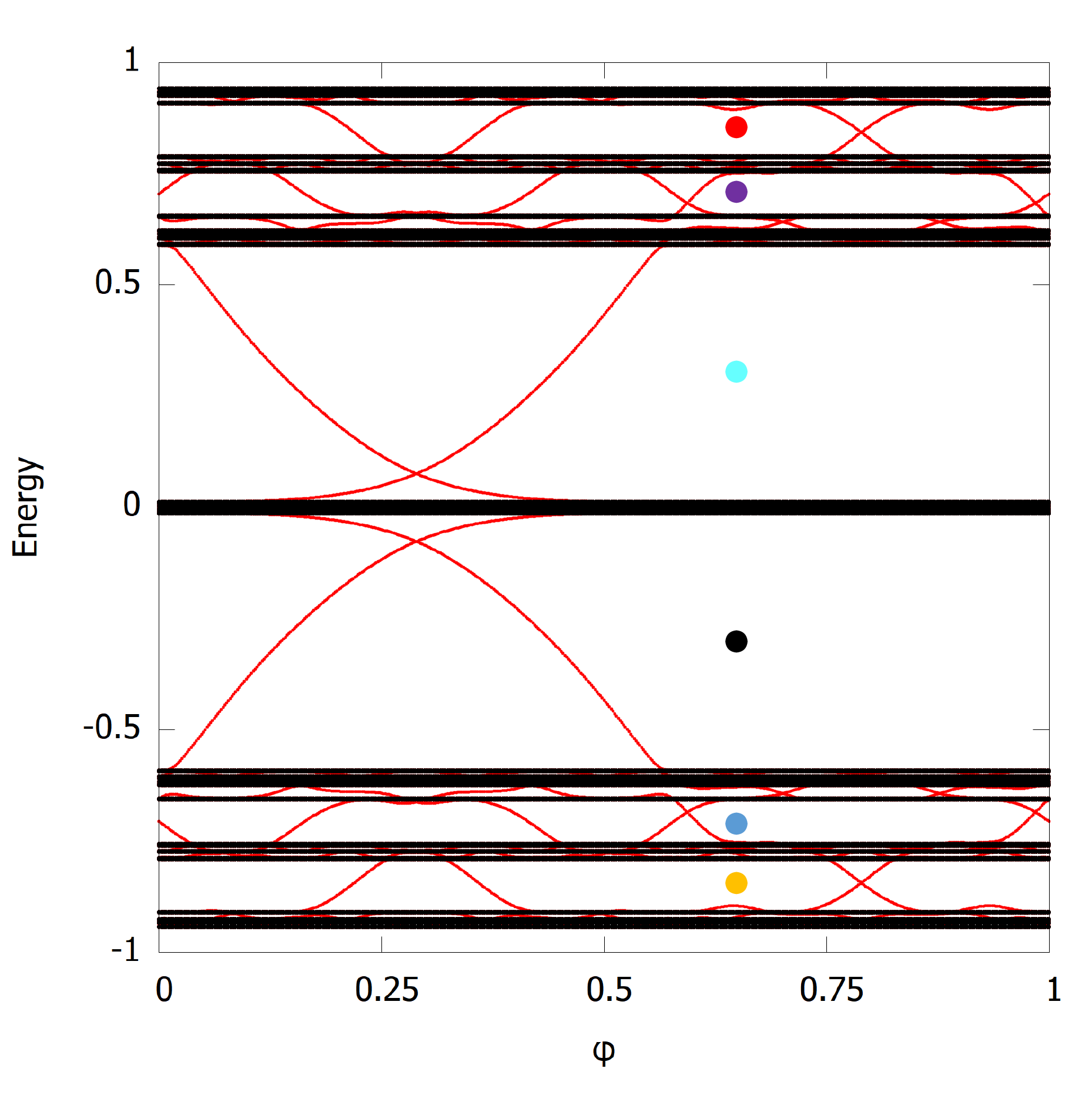}
		\caption{\small Energy spectrum of Hamiltonian \eqref{Eq:GenHam1} with open boundary condition inside the $M=1$ sector, plotted as function of parameter $\varphi$ from \eqref{Eq:PattAlg}. The topological boundary spectrum is highlighted in red and the spectrum with closed boundary conditions is overlaid in black color. The computation was performed with $J_z=0$, $|\Ll|=362$ and $\theta=\frac{3-\sqrt{3}}{3}$ specified in Fig.~\ref{Fig:M1JZ0Spec}. The colored dots mark the same gaps as in Fig.~\ref{Fig:M1JZ0Spec}.}
		\label{Fig:M1JZ0Phi}
	\end{figure} 

From the start, we consider here an infinite chain and view the Hamiltonians as unbounded self-adjoint operators on the Hilbert space $\Hh$. We now enlarge the class of models and to each finite system of continuous functions  $f_q : \SM \rightarrow \CM$, $q \in \ZM$, we associate a family of covariant Hamiltonians 
\begin{align}\label{Eq:GenHam4}
	H(\varphi) =&   \sum_q \sum_{n \in \ZM}  f_q(\varphi + n\theta) S^{+}_{n+q} S^{-}_{n} \\ \nonumber
	& \qquad \quad +J_z \sum_{n \in \ZM}   (S^{z}_n + \tfrac{1}{2})( S^z_{n+1} + \tfrac{1}{2}) - \mu M_z,
\end{align}
where $H$ is assumed self-adjoint. 	If all $f_q$'s are zero except for
\begin{equation}\label{Eq:F}
	f_1(\varphi) = \exp\Big [-\Big (\big |1+ r(\sin[2\pi(\varphi + \theta)]-\sin[2\pi \varphi])\big |\Big)\Big ],
	\end{equation}
then $J(d_n) = f(\varphi + n\theta)$ and the Hamiltonian \eqref{Eq:GenHam4} reduce to particular form in Eq.~\ref{Eq:GenHam1}. If $T_a$ with $a\in \ZM$ represent the usual translations of the spins, $T_a\vec S_n T_a^\dagger = \vec S_{n+a}$, from the new expression \eqref{Eq:GenHam4}, it is straightforward to derive the following covariance relation
\begin{equation}\label{Eq:CovariantR}
	T_a^\dagger H_S(\varphi) T_a = H_S(\varphi + a\theta).
\end{equation}
One of its direct implications is that  $H_S(\varphi + a\theta)$ all have identical spectra. To say more about these spectra, we need some knowledge about the continuity of the spectra with $\varphi$.

Using the estimates from \cite{BratelliBook2}[p.~249], one finds that the evolution operator $e^{-\imath t H_f(\varphi)}$ on $\Hh$ is norm continuous of $\varphi$ and $f_q$'s, for all $t \in \RM$. It then follows from \cite{OliveiraBook}[Th.~10.1.16] that $H(\varphi)$ is a continuous family w.r.t. both $f$ and $\varphi$ in the norm resolvent sense. Now, since the spectra are invariant under unitary transformations, it follows that ${\rm Spec}(H(\varphi+a\theta))$ are all the same for all $a\in \ZM$. When $\theta$ takes an irrational value, and only in that case, the orbit $\{(\varphi+a\theta){\rm mod}\, 1, \ a\in \ZM\}$ fills the circle $\SM$ densely. These facts together with the continuity w.r.t. $\varphi$ tell us that the spectrum of $H(\varphi)$ is completely independent of $\varphi$ when $\theta$ is an irrational number. Note that the covariance relation breaks down in the presence of a boundary, hence the boundary spectrum usually display dispersion w.r.t. $\varphi$.

	\section{The $M =1$ Sector}
	\label{Sec:M1Jz0}
	
\begin{figure}[t!]
		\includegraphics[width=\linewidth]{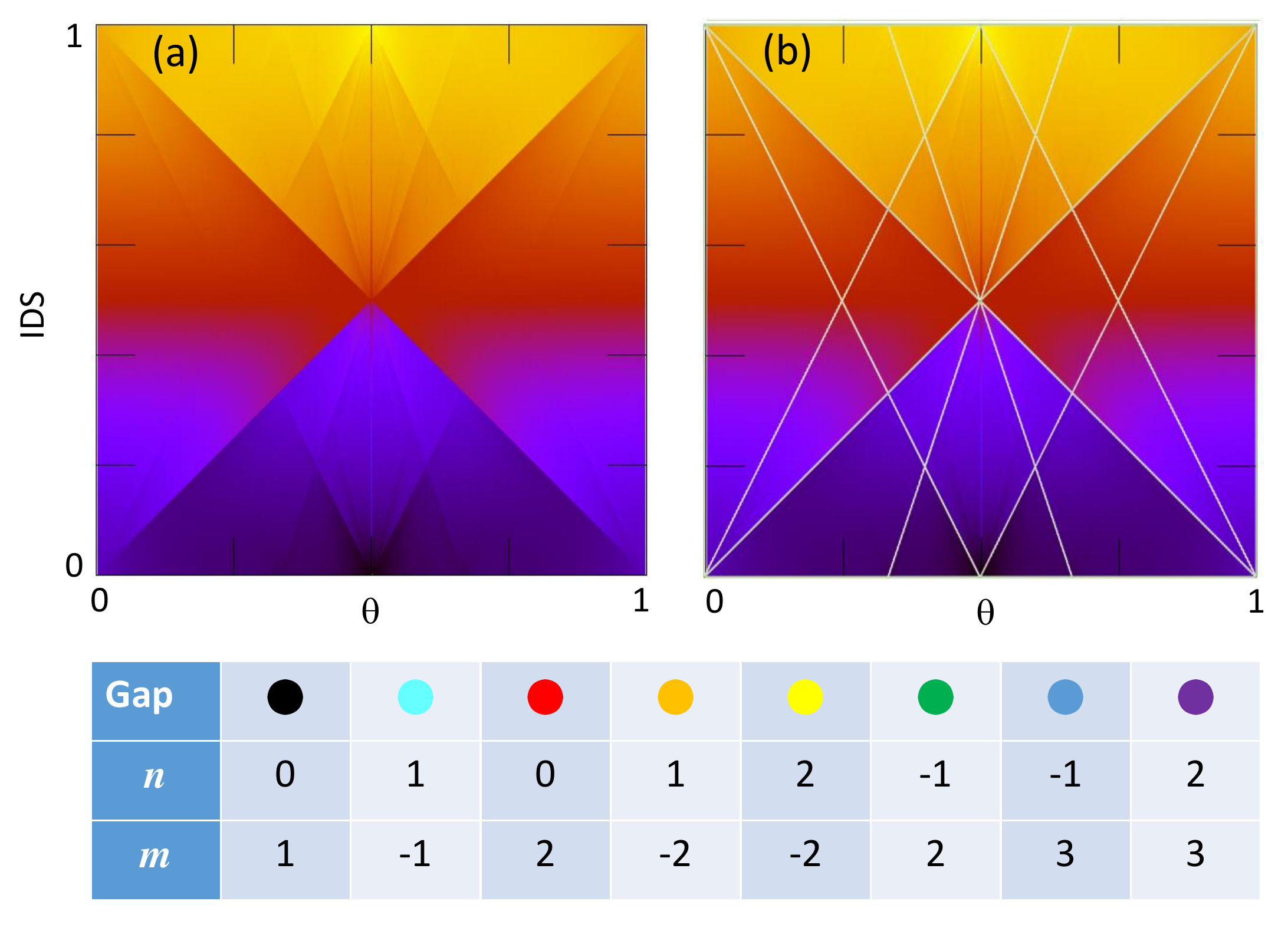}
		\caption{\small Numerical IDS as computed from the spectral butterfly reported in Fig.~\ref{Fig:M1JZ0Spec} for the $M=1$ sector and $J_z=0$. The predicted IDS values from Eq.~\eqref{Eq:IDS1Pred}, shown as light colored lines, are matched with the numerical IDS values inside the gaps marked in Fig.~\ref{Fig:M1JZ0Spec}, identified here by the abrupt changes in the color plot. The tables list the gap labels $(n,m)$ from Eq.~\eqref{Eq:IDS1Pred} as well as the corresponding gaps.}
		\label{Fig:GapLabels_M1JZ0}
	\end{figure}
	
	In this section, we briefly reproduce the analysis from \cite{LiuArxiv2020}, when the models defined in Eq.~\eqref{Eq:GenHam4} are restricted to the $M=1$ sector. The findings of this section will enable us to construct the gapped topological phases at finite magnetization densities.
	
In the $M=1$ sector, the interacting term drops out, hence we can set $J_z=0$ without altering the conclusions.	We start by reporting in Fig.~\ref{Fig:M1JZ0Spec} the bulk spectra of the Hamiltonian~\eqref{Eq:GenHam1} as a function of the parameter $\theta$. As one can see, they resembles quite closely the Hofstadter spectra of the electrons in a magnetic field \cite{Hoftadter1976}. In Fig.~
	\ref{Fig:M1JZ0Phi}, we report the spectrum of the same Hamiltonian computed with open boundary conditions at a fixed $\theta$ and variable parameter $\varphi$. Topological chiral edge modes can be observed in this plot. A brief explanation of these findings is supplied below. For more details, we point the reader to \cite{LiuArxiv2020}.
	
	\begin{figure}[t!]
		\includegraphics[width=\linewidth]{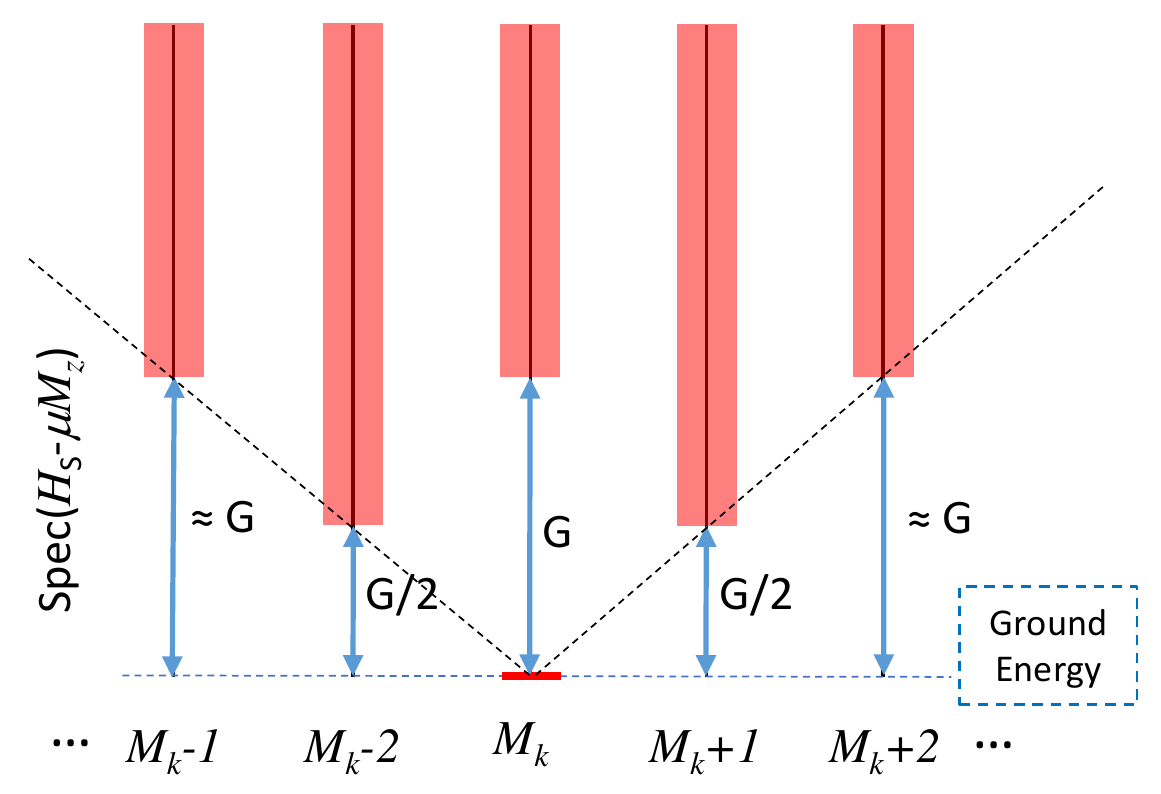}
		\caption{\small Schematic of the many-body spectrum at $J_z=0$ in different magnetization sectors, where $M_k$ is as in Eq.~\ref{Eq:MagicM}. A finite chain size is assumed, but the overall structure remains the same as the size of the chain is increased. The ground energy is shown by the red segment in the $M_k$ sector. The range of the excited energies are shown by the shaded regions. $G$ is the 1-particle spectral gap.}
\label{Fig:ManyBodySpectrum}
	\end{figure}
	
		\begin{figure*}[t!]
		\includegraphics[width=\linewidth]{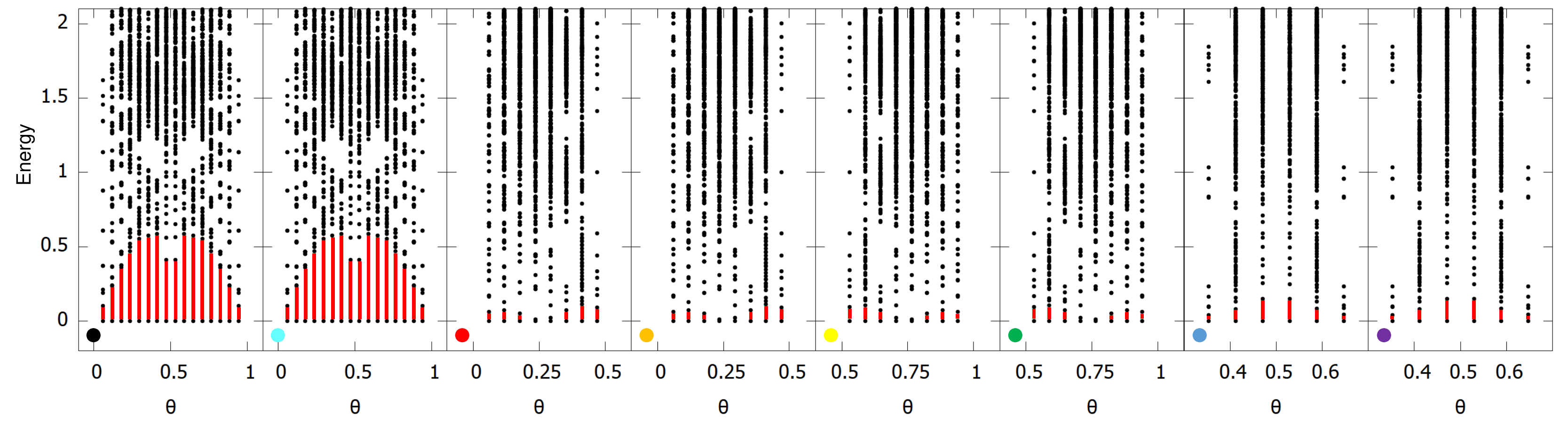}
		\caption{\small Energy spectra of $H_S$ as function of the quantized $\theta_k = k/|\Ll|$, computed inside the corresponding $M_k$ sectors singled out in Eq.~\eqref{Eq:MagicM}. The colored dots labeling the panels indicate the $M=1$ gaps in Fig.~\ref{Fig:GapLabels_M1JZ0} used to generate the gapped many-body states. The computation was performed with $J_z=0$ and $|\Ll|=17$. The energies are referenced from the bottom of the spectra in all panels and the many-body gaps are highlighted in red.}
		\label{Fig:SpectraL17Jz0}
\end{figure*}
	
	For the $M=1$ sector, a basis consists of states with all spins down and only one spin up. We denote the state with the up-spin located at  position $n$ by $|n\rangle$. The operators $S^+_{n+q} S^-_{n}$ act as simple hopping operators on these states and the spin models~\eqref{Eq:GenHam4} reduce to the ordinary tight-binding Hamiltonians
	\begin{equation}
	H_1 = \sum_q \sum_{n \in \mathbb Z}  f_q(\varphi+n\theta) \, |n+q\rangle \langle n| - \mu I.
	\end{equation}
Ignoring the trivial term $\mu I$, this expression, which also follows from the Jordan-Wigner transformation, can be re-written as
	\begin{align}\label{Eq:H1}
	H_1 = \sum_q S^q \sum_{n \in \mathbb Z}  f_q(\varphi+n\theta) \, |n\rangle \langle n|  
	\end{align}
where $S$ is the lattice shift operator $S|n\rangle = |n+1\rangle$. One can see now that all these models in the $M=1$ sector are generated by the shift operator $S$ and by diagonal operators of the form
	\begin{equation}\label{Eq:DiagOp}
	W_f=\sum_{n \in \mathbb Z}  f(\varphi+n\theta) \, |n\rangle \langle n|,
	\end{equation} 
	with $f$ a continuous function over the circle. Furthermore, we have the following commutation relation
\begin{equation}\label{Eq:Comm1}
	W_f \, S = S \, W_{f\circ \tau_\theta^1}.
	\end{equation}
	
\vspace{0.2cm}
	
	Any continuous function $f$ over the circle can be Fourier decomposed. As such, the algebra of continuous functions over the circle is generated by a single function,
	\begin{equation}
	v: \SM \rightarrow \CM, \quad v(x) = e^{\imath 2\pi x}.
	\end{equation} 
	Hence all the diagonal operators from Eq.~\eqref{Eq:DiagOp} can be obtained as linear combinations of powers of a single diagonal operator:
	\begin{equation}\label{Eq:U}
	V=e^{-\imath 2 \pi \varphi}\sum_{n \in \mathbb Z}  v(\varphi+n\theta) \, |n\rangle \langle n|=\sum_{n \in \mathbb Z}  e^{\imath 2 \pi n \theta} \, |n\rangle \langle n|.
	\end{equation}
	The conclusion is that all Hamiltonian $H_1$ can be drawn from the algebra $C^\ast(T,V)$ generated by $S$ and $V$, which obey the commutation relation
	\begin{equation}\label{Eq:Co1}
	VS = e^{\imath 2 \pi \theta} SV.
	\end{equation}
Hence, the algebra which generates all model Hamiltonians of type \eqref{Eq:GenHam1} coincides with the non-commutative 2-torus $\Aa_{\Theta}$ \cite{DavidsonBook}.
	
	\vspace{0.2cm}
	
	In Fig.~\ref{Fig:GapLabels_M1JZ0}, we report the integrated density of states (IDS) for Hamiltonian $H_1$, which has been directly computed from the spectrum ${\rm Spec}(H_1)$ reported in Fig.~\ref{Fig:M1JZ0Spec}, using the formula
	\begin{equation}\label{Eq:IDSN0}
	{\rm IDS}(E) = \frac{ \big |{\rm Spec}(H_1) \cap (-\infty,E] \big | }{|\Ll|}.
	\end{equation}
As in \cite{LiuArxiv2020}, the IDS is represented as a function of $\theta$ and energy, with the latter on the axis coming out of page. For visualization, the energy values are encoded in the color map and the striking features seen throughout this color map are the sudden changes in color, which occur along straight lines. These sudden changes in color correspond to the spectral gaps seen in Fig.~\ref{Fig:M1JZ0Spec} and the larger the gap the stronger the sudden change in color. This simple phenomenon enables us to determine the numerical values of the IDS inside the prominent spectral gaps marked in Fig.~\ref{Fig:M1JZ0Spec}. 

The IDS values inside a gap $G$ can be equivalently computed as
\begin{equation}
{\rm IDS}(G) = {\rm Tr}_{\rm L}(P_G),
\end{equation}
where $P_G$ is the gap projection onto the states below the gap $G$ and ${\rm Tr}_{\rm L}$ is the trace per length operation
\begin{equation}\label{Eq:TrL}
{\rm Tr}_{\rm L}\{\cdot\} = \lim_{|\Ll| \rightarrow \infty} \frac{{\rm Tr}\{ \cdot \}}{|\Ll|}.
\end{equation}
Since $P_G$'s are projections from the non-commutative 2-torus in the thermodynamic limit $|\Ll|\rightarrow \infty$, it is well known that IDS takes only the following quantized values inside the gaps \cite{FootnoteIDS}
	\begin{equation}\label{Eq:IDS1Pred}
	{\rm IDS}(G) \in \big \{ n + m \theta, \ n,m \in \ZM\big \} \cap [0,1].
	\end{equation}
The prediction is that IDS, when evaluated inside the gaps, displays only linear dependencies w.r.t. $\theta$ with integer coefficients. Fig.~\ref{Fig:GapLabels_M1JZ0} confirms that every single feature seen in the numerically computed IDS can be explained and matched by the prediction in Eq.~\eqref{Eq:IDS1Pred}. The resulting $(n,m)$-labels are reported in the table in Fig.~\ref{Fig:GapLabels_M1JZ0} for all the marked gaps in Fig.~\ref{Fig:M1JZ0Spec}. 

\begin{figure*}[t!]
\includegraphics[width=\linewidth]{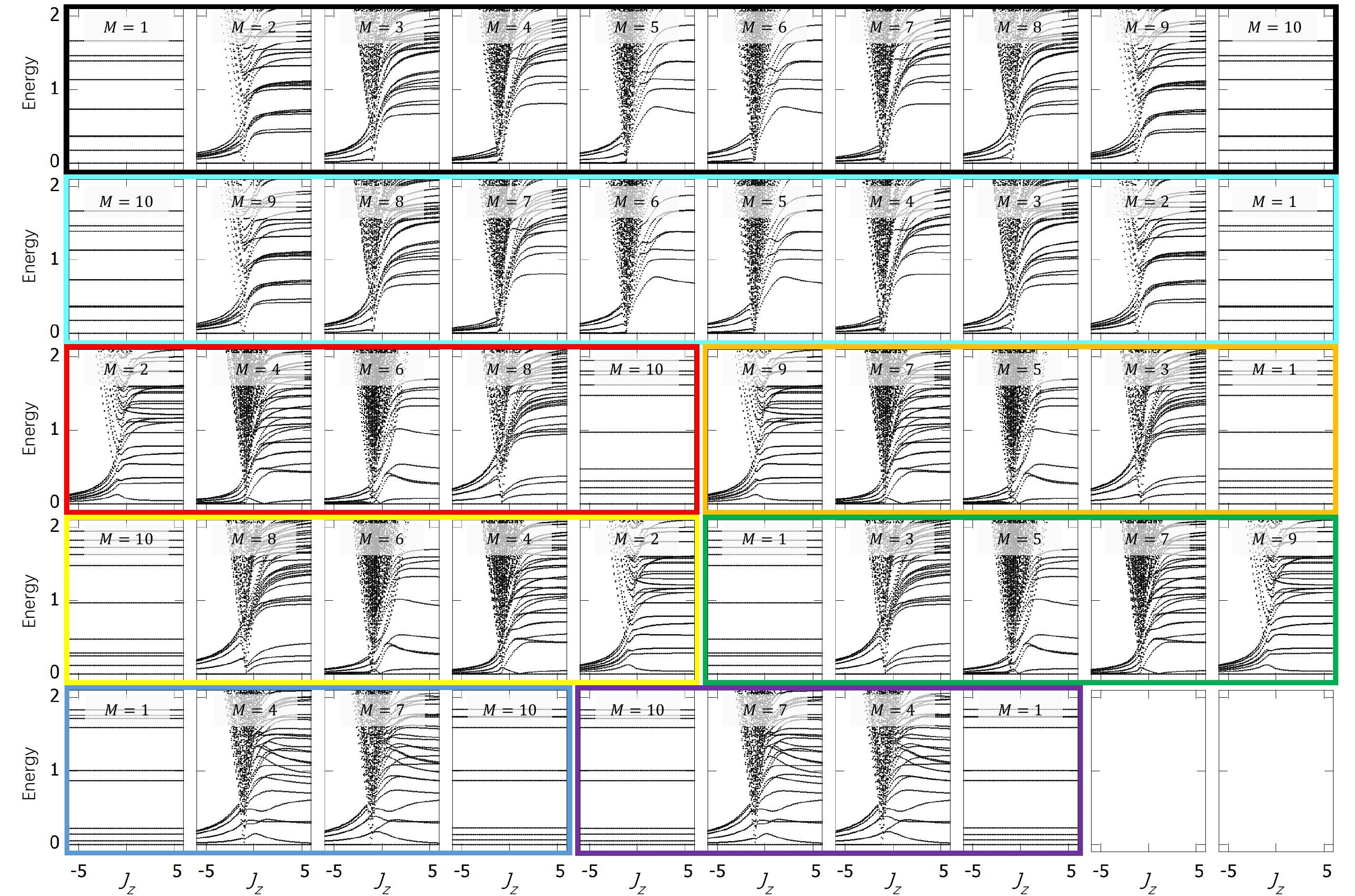}
\caption{\small Dependence of the energy spectra of the Hamiltonian \eqref{Eq:GenHam1} on $J_z$ for different $M_k$ sectors, with $M_k$ as in \eqref{Eq:MagicM}. The colors of the frames match the colors of the markers in Figs.~\ref{Fig:M1JZ0Spec} and \ref{Fig:GapLabels_M1JZ0}. They are used to indicate the $(n,m)$ gap labels appearing in \eqref{Eq:MagicM} and also the $M=1$ gaps in Fig.~\ref{Fig:M1JZ0Spec} used to generate the many-body states in each panel. The computation was performed with $|\Ll|=11$. The energies are referenced from the bottom of the spectra in all panels.}
\label{Fig:SpectraL11Jz5}
\end{figure*}

We recall that the $m$-coefficient coincides with the first Chern number of the gap projection, which follows directly from the Streda formula \cite{StredaJPC1982}. As such, any gap carrying a non-zero $m$-label should display $m$ topological edge modes when the boundary condition is switched from closed to open. This bulk-boundary correspondence is well understood \cite{ProdanSpringer2016} and it is indeed confirmed by Fig.~\ref{Fig:M1JZ0Phi} and the gap labels mapped in Fig.~\ref{Fig:GapLabels_M1JZ0}. The simulations in Fig.~\ref{Fig:M1JZ0Phi} were carried with open boundary conditions, hence, the chain displays two edges and this is why the number of topological edge bands are doubled in Fig.~\ref{Fig:M1JZ0Phi}.
	
\section{Many-Body Topological Gaps}
\label{Sec:MBodyGaps}
	
In this section, we use the results from $M=1$ sector to construct new gapped many-body phases at finite magnetization densities 
\begin{equation}
\mathfrak m = \lim_{|\Ll|\rightarrow \infty} \frac{M}{|\Ll|} >0.
\end{equation}
	
We start from the un-correlated case $J_z=0$ and use the fermionic representation $H_F$ of the model. Then Fig.~\ref{Fig:M1JZ0Spec} can be interpreted as the 1-particle spectrum and one knows that gapped many-particle states are obtained if and only if all the states below a 1-particle gap are fully populated. Given the correspondence in Eq.~\eqref{Eq:MtoN}, we conclude that, within the sectors with \begin{equation}\label{Eq:MTheta}
\mathfrak m_\theta = n + m \theta,
\end{equation} 
$H_S$ should display a non-degenerate ground level, separated from the rest of its spectrum by a gap that can be read off from Fig.~\ref{Fig:M1JZ0Spec}.

For finite chains, this led us to investigate the sectors $M = (n + m \theta) |\Ll|$. We recall that, in order to comply with the closed boundary conditions, the parameter $\theta$ is always quantized as $\theta_k = k /|\Ll|$, $k = 1, \ldots, |\Ll|$. As such, in our finite size simulations, we probed the magnetization sectors
\begin{equation}\label{Eq:MagicM}
M_k = n |\Ll| + m k,
\end{equation}
where only the $k$'s for which $0 \leq M_k \leq \Ll$ are allowed. In Fig.~\ref{Fig:ManyBodySpectrum}, we sketch the spectrum in different $M$ sectors at fixed $\theta_k$, based on the fermionic representation \eqref{Eq:FHam} and on populating the 1-particle spectrum shown in Fig.~\ref{Fig:M1JZ0Spec} with appropriate number of fermions. The structure of the spectrum seen in Fig.~\ref{Fig:ManyBodySpectrum} will play a major role in the discussion below. The calculated spectra of the Hamiltonian~\eqref{Eq:GenHam1} (with $J_z=0$) inside the sectors \eqref{Eq:MagicM} for all allowed $k$'s and for different $(n,m)$ gap labels are reported in Fig.~\ref{Fig:SpectraL17Jz0}. The results confirm the non-degenerate character of the ground states for all the cases where the single particle gaps are opened in Fig.~\ref{Fig:M1JZ0Spec}.

	\begin{figure*}[t!]
		\includegraphics[width=0.8\linewidth]{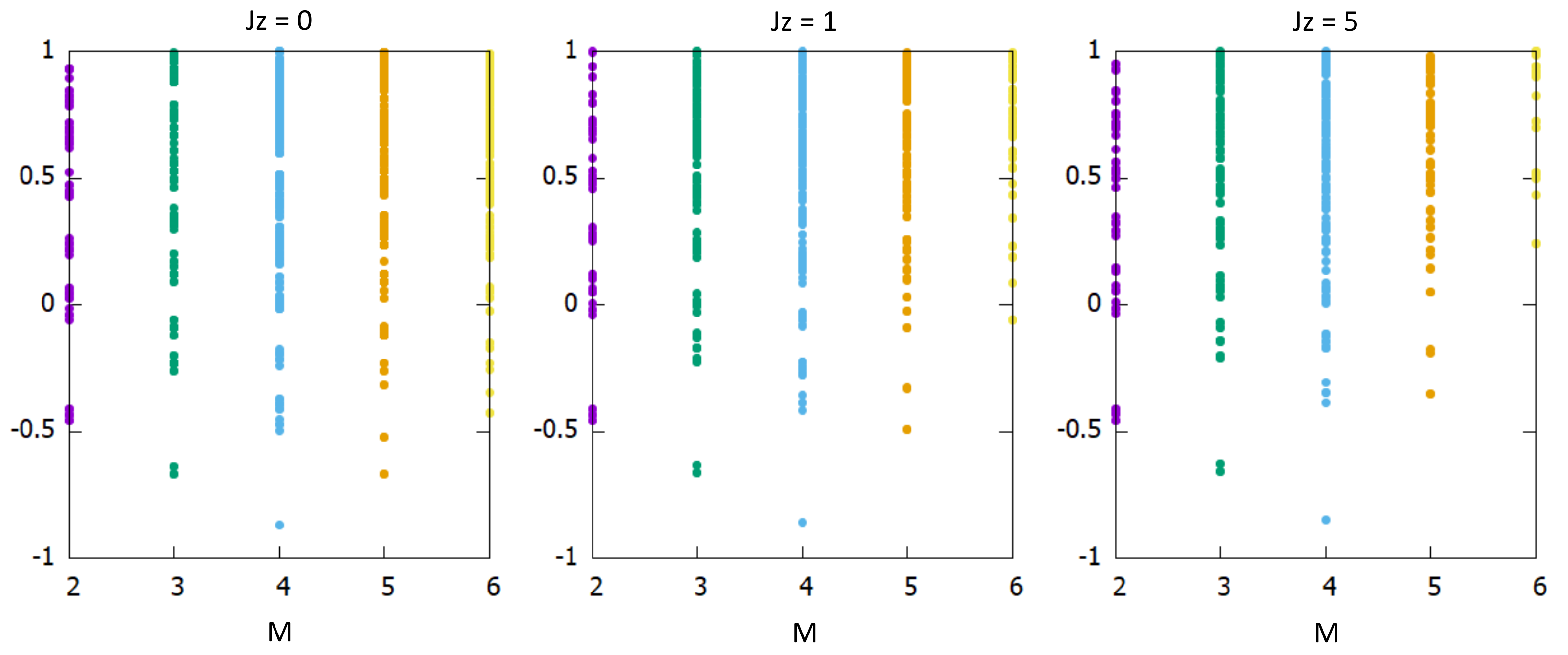}
		\caption{\small Evolution of the spectrum of the Hamiltonian~\eqref{Eq:GenHam1} with $J_z$ inside various $M$ sectors. The calculation was performed for $|\Ll|=17$ and $M=4$ represents the special $M_k$ in Eq.~\eqref{Eq:MagicM} for the gap labels $(0,1)$ corresponding to the \color{black}{$\bullet$}-gap in Fig.~\ref{Fig:M1JZ0Spec}.}
		\label{Fig:MManyBodySpec}
\end{figure*}

The flows of the bulk spectra w.r.t. the interaction parameter $J_z$ inside the range $[-5,5]$ are reported in Fig.~\ref{Fig:SpectraL11Jz5}. As one can see, all many-body ground state gaps in the appropriate magnetization sectors remain open for all $J_z$ values, except for several distinct values where the many-body gaps close. However, an investigation of the spectra for increasing chain sizes reveal that the small gaps to the left of those gap closings are unstable and seem to close as the size of the system is increased to infinity. As such, we will primarily focus on the many-body gaps seen to the right of those transition points. 

We recall that $H_s$ is defined on the full Hilbert $\bigoplus_{M} \Hh_M$ and, as such, its full energy spectrum is supplied by the union of all the spectra corresponding to different magnetization sectors. We then focus again at the spectrum sketch in Fig.~\ref{Fig:ManyBodySpectrum} for $J_z=0$ and point out that the lowest non-degenerate eigenvalue $E_0$ detected inside the $M_k$ sector is in fact the lowest eigenvalue in the full spectrum, hence the ground state for $H_S(\varphi)$. The structure of the spectrum seen in Fig.~\ref{Fig:ManyBodySpectrum} is independent of the chain's length, but note that $E_0 \rightarrow - \infty$ as the length of the chain is increased. In Fig.~\ref{Fig:MManyBodySpec}, we present evidence that same generic picture holds for $J_z >0$. We have also investigated, as far as we could, that the structure of the spectrum at $J_z >0$ remains stable against the chain's size.

\begin{table}[b!]
		\centering
		\scriptsize
		\begin{tabular}{|c|c|c|c|c|}
			\hline
			\multicolumn{1}{|c|}{gap \& labels} & \multicolumn{2}{c|}{parameters}  & \multicolumn{2}{c|}{results} \\
			\hline
			$(n,m)$	&$M_k$ & $\theta_k$ & $\Delta E$& ${\rm Ch}(P)$   \\
			\hline
			\color{black}{\large $\bullet$} \ (0,1)& 1&0.09&0.30& 1.00000000000000\\
			& 2&0.18&0.63& 1.00000000000000\\
			& 3&0.27&0.89& 1.00000000000000\\
			& 4&0.36&1.00& 1.00000000000000\\
			& 5&0.45&0.84& 0.999999999999999\\
			& 6&0.55&0.84& 0.999999999999999\\
			& 7&0.64&1.00& 1.00000000000000\\
			& 8&0.73&0.89& 1.00000000000000\\
			& 9&0.82&0.63& 1.00000000000000\\
			&10&0.91&0.30& 1.00000000000000\\
			\hline
			\color{red}{\large $\bullet$} \ (0,2)& 2&0.09&0.56& 2.00000000000000\\
			& 4&0.18&0.36& 2.00000000000000\\
			& 6&0.27&0.11& 1.99999999999999\\
			& 8&0.36&0.64& 2.00000000000000\\
			&10&0.45&1.00& 2.00000000000000\\
			\hline
			\color{bleudefrance}{\large $\bullet$} \ (-1,3)& 1&0.36&0.30& 3.00000000000000\\
			& 4&0.45&0.89& 3.00000000000000\\
			& 7&0.55&1.00& 3.00000000000000\\
			&10&0.64&0.37& 3.00000000000000\\
			\hline
		\end{tabular}
		\caption{\small Comparison between the gap labels used to generate the many-body gapped states and the numerically evaluated Chern numbers \eqref{Eq:Chern} at $J_z=0$. This calculation was performed with $|\Ll|=11$. $\Delta E$ is the size of the many-body gap normalized to its maximum value when varying $M$.}
		\label{Table:ChL11Jz0}
	\end{table}

	\begin{table}[b!]
		\centering
		\scriptsize
		\begin{tabular}{|c|c|c|c|c|}
			\hline
			\multicolumn{1}{|c|}{gap \& labels} & \multicolumn{2}{c|}{parameters}  & \multicolumn{2}{c|}{results} \\
			\hline
			$(n,m)$	&$M_k$ & $\theta_k$ & $\Delta E$& ${\rm Ch}(P)$   \\
			\hline 
			\color{black}{\large $\bullet$} \ (0,1)& 1&0.08&0.24&0.999999999999998\\
			& 2&0.15&0.53&0.999999999999998\\
			& 3&0.23&0.77&0.999999999999999\\
			& 4&0.31&0.94&0.999999999999999\\
			& 5&0.38&1.00&1.00000000000000 \\
			& 6&0.46&0.77&1.00000000000000 \\
			& 7&0.54&0.77&1.00000000000000 \\
			& 8&0.62&1.00&0.999999999999999\\
			& 9&0.69&0.94&0.999999999999999\\
			&10&0.77&0.77&0.999999999999999\\
			&11&0.85&0.53&1.00000000000000 \\
			&12&0.92&0.24&1.00000000000000 \\
			\hline
			\color{red}{\large $\bullet$} \ (0,2) &2&0.08&0.64&2.00000000000000\\
			& 4&0.15&0.57&2.00000000000000\\
			& 6&0.23&0.12&1.99999999999997\\
			& 8&0.31&0.23&2.00000000000000\\
			&10&0.38&1.00&2.00000000000000\\
			&12&0.46&0.98&2.00000000000000\\
			\hline
			\color{bleudefrance}{\large $\bullet$} \ (-1,3)& 2&0.38&0.36& 3.00000000000000\\
			& 5&0.46&0.95& 3.00000000000000\\
			& 8&0.54&1.00& 3.00000000000000\\
			&11&0.62&0.45& 3.00000000000000\\
			\hline
		\end{tabular}
		\caption{\small Same as Table~\ref{Table:ChL11Jz0} but for $|\Ll|=13$.}
		\label{Table:ChL13Jz0}
	\end{table}

The above findings support the conclusion that Gibbs states 
\begin{equation}\label{Eq:GState}
\omega_G(A) = \lim_{\beta \rightarrow \infty}\lim_{|\Ll| \rightarrow \infty} \frac{{\rm Tr}\{A e^{-\beta H_S}\}}{{\rm Tr}\{e^{-\beta H_S}\}}
\end{equation}
defines a true ground state on the algebra of local spin operators. We will come back to this state and its GNS representation in Sec.~\ref{Sec:Interpretation}.
	
	\section{Mapping the Chern numbers}
\label{Sec:Chern}

\begin{figure*}[t!]
\includegraphics[width=\linewidth]{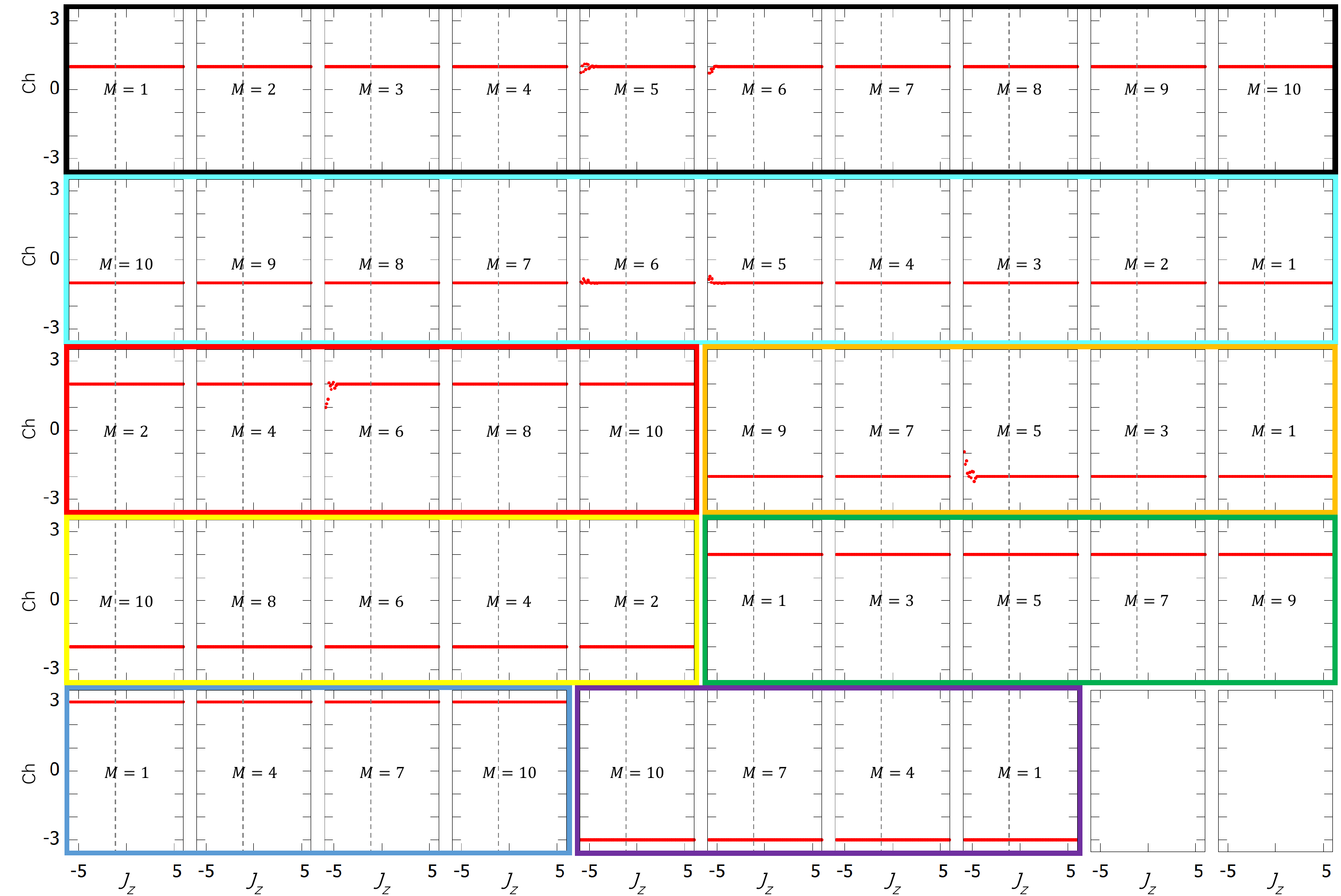}
\caption{\small The Chern number \eqref{Eq:Chern} as function of $J_z$ for the gapped many-body states shown in Fig.~\ref{Fig:SpectraL11Jz5}. The color scheme and the labelings are as in Fig.~\ref{Fig:SpectraL11Jz5}. The vertical dotted lines indicate the gap closings observed in Fig.~\ref{Fig:SpectraL11Jz5}.}
\label{Fig:ChernL11Jz5}
\end{figure*}

One standard way to generate topological invariants is through torus actions \cite{CareyAMS2014}[Sec.~5.1] on the algebra of physical observables. Recall that the circle $\SM = \RM/\ZM$ comes equipped with a natural group structure and so does the 2-torus $\TM = \SM \times \SM$. A torus action on an algebra is simply a group homomorphism from $\TM^2$ to the group of automorphism of that algebra. 

On the algebra generated by our spin models, we already have a circle action given by the shift $\varphi \mapsto (\varphi+a){\rm mod}\, 1$ of the parameter in Eq.~\eqref{Eq:PattAlg}. We construct another circle action via the transformations
\begin{equation}\label{Eq:TorusAction}
\left \{
	\begin{aligned}
	\rho_w(S_n^+) & = e^{\imath 2\pi nw}S_n^+,\\ 
	\rho_w(S_n^-) & = e^{-\imath 2\pi nw}S_n^-, \\ 
	\rho_w(S^z_n) & = S_n^z,
	\end{aligned}
	\right .
	\end{equation}
of the generators. It is straightforward to verify that $\rho_w$ preserves the commutation relations for the spin operators. As such, $\rho_w$ extends to a full automorphism on the algebra of local spin observables and, furthermore, $\rho_{w+w'} = \rho_w \, \rho_{w'}$. As a result, we obtained the second circle action, needed for a full torus action. 

Before supplying the expression for the Chern number, we discuss the physics of the $w$-action. First, we note that the generator of the $w$-transformation is
\begin{equation}
\frac{1}{2 \pi}\left . \frac{{\rm d} \rho_w(S_n^\pm)}{{\rm d} w}\right |_{w=0} = \imath [X,S_n^\pm], \quad X = \sum_{n \in \ZM} n (S_n^z+\tfrac{1}{2}).
\end{equation}
Under the Jordan-Wigner mapping, this observable becomes the position operator $X = \sum_{n \in \ZM} n a_n^\ast a_n$. In the fermionic representation, one will also find from Eq.~\ref{Eq:StoA} that $\rho_w(a_n) = e^{-\imath 2 \pi n w} a_n$, which is the standard $U(1)$-gauge tranformation related to the charge transport. For example, the charge current operator for the fermionic model is just ${\rm d}\rho_w(H_F)/{\rm d}w$. Furthermore, if we pass to the Fourier transformed operators, then $\rho_w(a_k) = a_{k+w}$. As such, $\rho_w$ is just a shift on the 1-particle Brilloine torus. There is also a connection with the twisted boundary conditions \cite{ShengPRL2003,WanPRB2005,KudoPRL2019}. Indeed, working with the fermionic representation of the models, we were able to derive that, for a finite size chain, the twisted boundary condition by $e^{\imath 2 \pi w}$ is implemented by $\rho_{w/|\Ll|}$. Note, however, that this makes little sense in the thermodynamic limit.

Now, the torus action \eqref{Eq:TorusAction} supplies a family of Hamiltonians $H_S(\varphi,w)=\rho_w(H_S(\varphi))$ indexed by $\TM^2$, which take the explicit form
\begin{equation}
\begin{aligned}
	H_S(\varphi,w)=\sum_n & \Big [J(d_n) \big (e^{-\imath 2\pi w}S_n^+S_{n+1}^-+e^{\imath 2\pi w}S_n^-S_{n+1}^+ \big )\\ 
	 & \quad +J_zS_n^zS_{n+1}^z \Big].
\end{aligned}
\end{equation}
Let us recall that the spectrum of $H_S(\varphi,w)$ is independent of both $\varphi$ and $w$. As such, the gapped states identified in the previous section supply a family of projections $P(\varphi,w)$ parametrized by the 2-torus, projecting on the appropriate magnetization sectors $\mathfrak m_\theta$ and on the ground levels of $H_S(\varphi,w)$. We then define the Chern number as
\begin{align}\label{Eq:Chern}
	\text{Ch}(P)=\tfrac{\imath}{2\pi}\int_{\TM^2} \text{d} \varphi \text{d} w\ {\rm Tr}_{\rm L}\left\{P \left[\frac{\partial P}{\partial \varphi},\frac{\partial P }{\partial w}\right]\right\}.
\end{align}
where ${\rm Tr}_{\rm L}$ is as in \eqref{Eq:TrL} but on the full representation space $\Hh$. Obviously, ${\rm Tr}_{\rm L}$ is a trace state which is invariant to the torus action. Furthermore, we have ${\rm Tr}_{\rm L}\{M_z P_\theta\} = \mathfrak m_\theta$.

\begin{figure}[t!]
		\includegraphics[width=\linewidth]{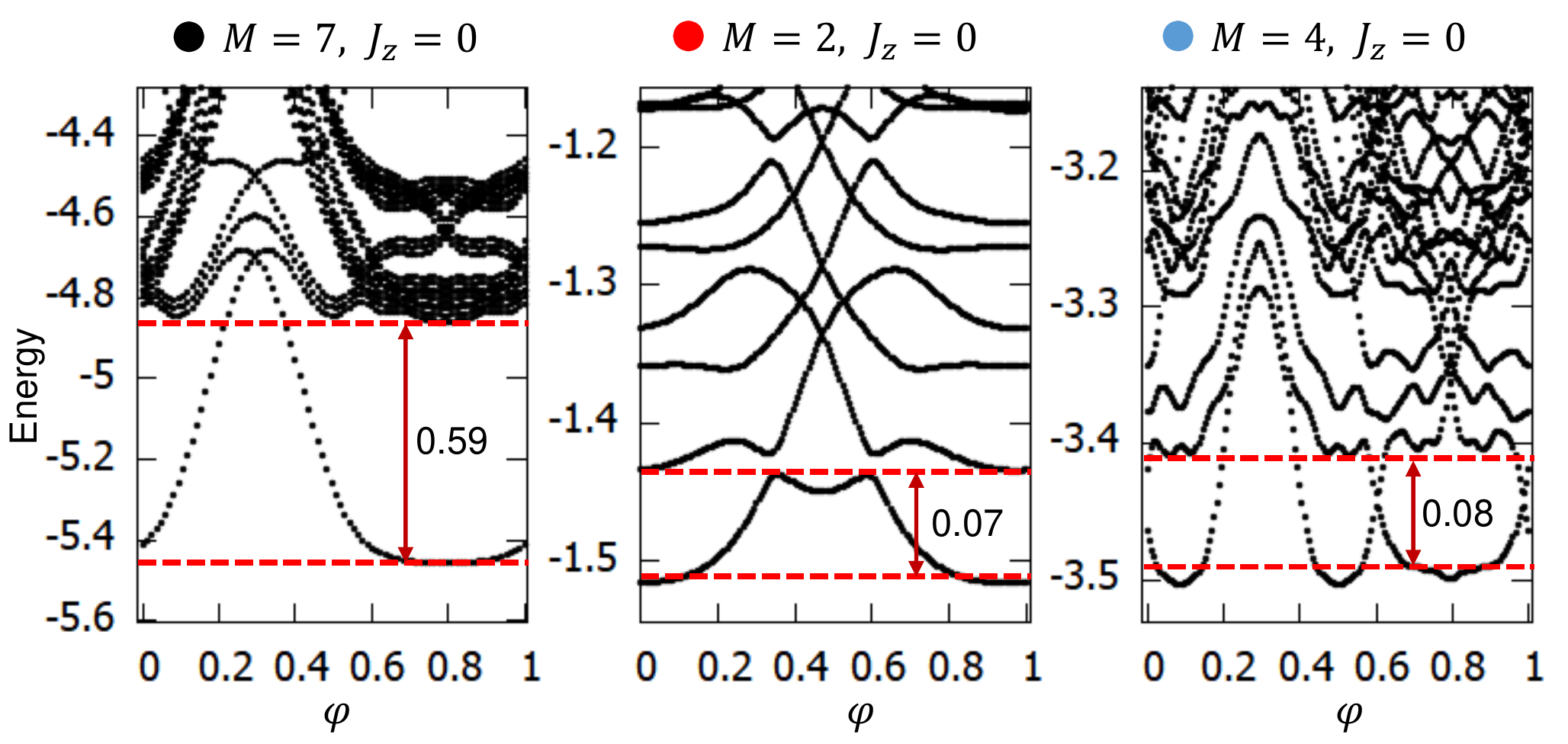}
		\caption{\small Energy spectra of $H_S$ ($J_z=0$), computed with open boundary conditions at fixed $\theta_k$ and inside the corresponding sector $M_k$ (shown for each panel). The spectra are shown as functions of parameter $\varphi$. The computation was performed with $|\Ll|=17$. The energies are referenced from the bottom of the spectra in all panels. The colored dots mark the same gaps as in Fig.~\ref{Fig:GapLabels_M1JZ0}.   }
\label{Fig:EdgeSpecJz0}
\end{figure}

The expression \eqref{Eq:Chern} will be numerically evaluated using finite chains. To comply with the closed boundary conditions, we need to restrict $w$ to quantized values $w_k = k/|\Ll|$. An identical quantization will be also considered for $\varphi$. This results in a discretized torus and we adopt the efficient algorithm from \cite{Fukui2005} to evaluate the Chern number from such discretized tori. Let us recall that, for a vector bundle over a 2-torus and with the ordinary trace replacing our ${\rm Tr}_{\rm L}$ operation, the algorithm in \cite{Fukui2005} always returns an integer number. This integer number may fluctuate for coarser discretizations of the torus but it stabilizes to the true values for finer discretizations. For example, a finer discretization will be needed if the system is close to a topological phase transition.

We want to point out that our formula differs fundamentally from the ordinary Chern formula because the ordinary trace is replaced by the trace per length. Then, in the thermodynamic limit, the expression \eqref{Eq:Chern} can, in principle, return any value from the real axis. However, in Tables~\ref{Table:ChL11Jz0} and \ref{Table:ChL13Jz0}, we report the numerical evaluation of Eq.~\eqref{Eq:Chern}, for $J_z=0$  and chain sizes $|\Ll|=11$ and $13$. As one can see, there are no fluctuations in the data and the numerical outputs correlate perfectly with the 1-particle gap labels. In fact, for $J_z=0$, it can be shown explicitly that the expression \eqref{Eq:Chern} reduces to that of the ordinary Chern number for the corresponding gap projection in the $M=1$.

In Fig.~\ref{Fig:ChernL11Jz5}, we report on the stability of the Chern numbers with respect to the interaction strength. As one can see, the values remain quantized even for negative $J_z$ values where a many-body gap closes and opens again. This can be seen as a reflection of Theorem~1 from \cite{LuAOP2020}, which says that the values of the Chern numbers are pinned by the filling factors, which in turn are fixed by the gap labels. However, we have already mentioned that the gaps at negative $J_z$ are unstable in the thermodynamic limit, hence one should only thrust the results in Fig.~\ref{Fig:ChernL11Jz5} at positive $J_z$'s.

The connection with the work in \cite{LuAOP2020} can be established quite explicitly. Indeed, the operators $e^{\imath 2 \pi \theta X}$ and the shift operator $T_1$ of the spins obey the commutation relations
\begin{equation}\label{Eq:CommX}
e^{\imath 2 \pi \theta X} T_1 = e^{\imath 2 \pi \theta M_z}T_1 e^{\imath 2 \pi \theta X}.
\end{equation}
Our models commute with the ``magnetic translations'' dual to the above (see \cite{ProdanSpringer2016}[p.~24]). This dual algebra is explicitly generated by the operators
\begin{equation}
V_1|\varphi,\{s\}\rangle = e^{\imath 2 \pi \varphi M_z}|\varphi,\{s\}\rangle, \quad V_2|\varphi,\{s\}\rangle = |\varphi+\theta,\{s\}\rangle,
\end{equation}
where $\{s\}$ is a spin configuration and the dependency on $\varphi$ of the states has been explicitly inserted into the notation. It is straightforward to check that $V_i$'s obey a commutation relation as in \eqref{Eq:CommX} and that, indeed, the generic models \eqref{Eq:GenHam4} commute with any element from the algebra generated by $V_1$ and $V_2$. As such, we are indeed in the conditions of Theorem~1 from \cite{LuAOP2020}. As we already stated, our results fully confirm the statements from \cite{LuAOP2020}. 

\begin{figure}[t!]
		\includegraphics[width=\linewidth]{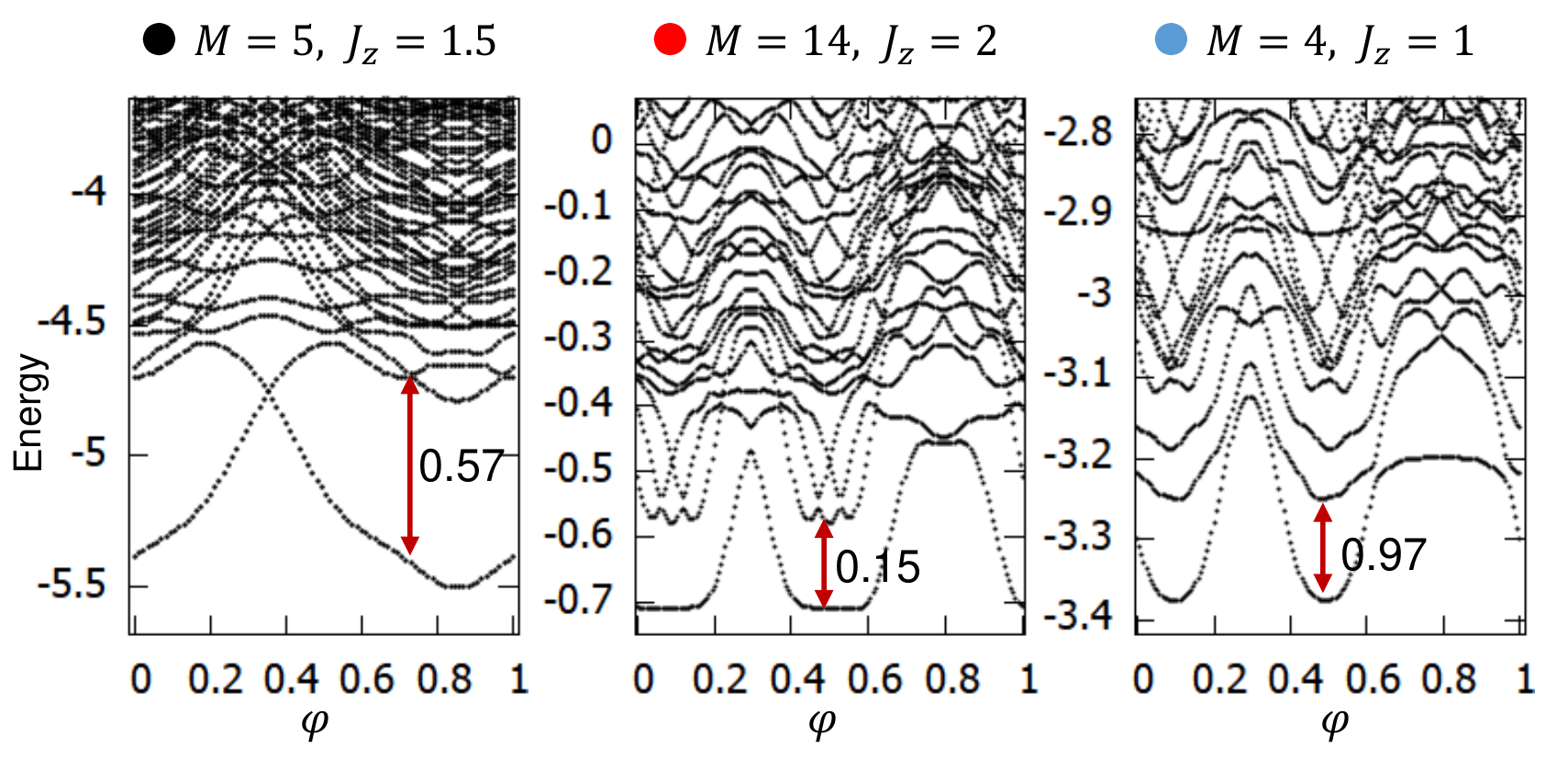}
		\caption{\small Same as Fig.~\ref{Fig:EdgeSpecJz0} but at finite $J_z$'s.}
		\label{Fig:PhiDependenceL17Jz}
\end{figure}

\section{Edge States}

When the boundary condition is switched from closed to opened, the 1-particle spectrum reported in Fig.~\ref{Fig:M1JZ0Phi} indicates to us that the many-body gaps reported in Fig.~\ref{Fig:SpectraL17Jz0} will close if we vary $\varphi$. This is confirmed in Fig.~\ref{Fig:EdgeSpecJz0} for several many-body gaps and we verified separately that this is the case for all the remaining gaps. Same observations hold at finite $J_z$'s, as seen in Fig.~\ref{Fig:PhiDependenceL17Jz}. 

The bulk-boundary correspondence in this many-body context presents several outstanding difficulties. First, we note that, even at $J_z=0$, the flow of the ground state energy with $\varphi$ is not entirely determined by the edge bands inside the $M=1$ gap that generated the many-body state in the first place. For example, all edge states emerged in the $M=1$ gaps below that gap will contribute to the ground state energy. For the same reason, the spatial profile of the ground state cannot be determined solely from the edge modes emerging in the $M=1$ gap that generated the many-body state in the first place. Secondly, examining only the magnetization sector where the gapped state was found may not provide the full picture of the edge physics. For example, if one edge is sent to infinity and the chemical potential is held fixed, then it is possible that the ground state of the system to reside in the neighboring magnetization sectors.  Lastly, it is not clear yet how to generalize the boundary topological invariant \cite{KellendonkRMP2002}, used to establish the bulk-boundary correspondence for the integer quantum Hall effect, to the the many-body context.

These challenges will be undertaken in our future investigations. For now, the only conclusion that emerges from our study is that the non-degenerate character of a ground state carrying a non-trivial Chern number is destroyed when open boundary conditions are used and $\varphi$ is varied over its entire range.
	
\section{Interpretation and Outlook}
\label{Sec:Interpretation}

There are two important aspects which were left un-answered in the previous section. One is the meaning of the ground state projection $P$ in the thermodynamic limit. One can try to find its meaning in the algebra $B(\Hh)$ of bounded operators over the Hilbert space $\Hh$, but the efforts will be futile. As a sequence of rank-1 projections, the finite size projections meander inside that algebra without a limit. If one uses the strong or weak topologies on $B(\Hh)$, one will find that this sequence of projections converges to zero. The other un-answered question is what is the algebra generated by our spin models? We recall that this question had a very sharp answer when we restricted the analysis to the $M=1$ sector. The discussion immediately below is technical and intended for the experts.

The two questions are in fact related and, to give tentative answers, we recall the ground states $\omega_G$ defined in Eq.~\eqref{Eq:GState} on the algebra of local spin observables $\Ss$, whose existence in the thermodynamic limit is assumed from now on. We recall that $\Ss$ has a precise definition in the thermodynamic limit as a $C^\ast$-algebra (see \cite{BratelliBook2}[Sec.~6.2]). These states, as any other states, supply GNS representations $\pi_G$ of the local spin algebra $\Ss$ (for more on GNS construction, see \cite{DavidsonBook}). Now, each GNS representation comes with a specific Hilbert space $\Hh_G$ and a cyclic vector $|\Psi_G\rangle$, such that $\omega_G(A) = \langle \Psi_G|\pi_G(A)|\Psi_G \rangle$ for each element $A \in \Ss$. If we take into account the $\varphi$ and $w$ dependencies, we then have a family $|\Psi_G(\varphi,w) \rangle$ of cyclic vectors indexed by the torus. Before claiming that $|\Psi_G(\varphi,w) \rangle$ supplies the vector bundle over the torus we were looking for, we must answer the following fundamental question: How do we relate these cyclic vectors for different values of the parameters or, more precisely, what is the topology that glues $|\Psi_G(\varphi,w) \rangle$ together? For answers, we examine more closely the GNS construction, which starts from the pre-Hilbertian scalar product $(A,B)_G = \omega_G(A^\ast B)$ on the algebra $\Ss$ itself. To transform $\{\Ss,(\cdot,\cdot)_G\}$ into a true Hilbert space, the $\Hh_G$ we already mentioned, we need to mod out the space of zero norm elements, that is, the null space
\begin{equation}
\Nn_G = \{A \in \Ss, \ \omega_G(A^\ast A)=0\},
\end{equation}
and complete the quotient space. Then the cyclic vector is just the class of unit element of $\Ss$ in this quotient space, that is,
\begin{equation}
|\Psi_G\rangle = |1 + \Nn_G\rangle.
\end{equation} 
One now can see that the link between various $|\Psi_G(\varphi,w) \rangle$ is supplied by the null spaces $\Nn_G(\varphi,w)$, which all live inside $\Ss$, hence they can be glued together by using the topology of $\Ss$.

To accomplish this last step, we recall that the null space $\Nn_G$ is in fact a closed left ideal of $\Ss$. As such, it is also a left (countably generated) $\Ss$-Hilbert module. Then, Theorem~1.3.5 in \cite{ThomsenBook} assures us that $\Nn_G = \PM_G \Hh_\Ss$, where $\Hh_\Ss$ is the standard Hilbert module over $\Ss$ (see \cite{ProdanRMP2016}[Example~3.18]) and $\PM_G$ is a projection from the algebra $B(\Hh_\Ss)$ of adjointable endomorphisms over $\Hh_\Ss$. We claim that this projection $\PM_G$, which is unique up to unitary conjugations, correctly encapsulates the thermodynamic limit of the finite size ground state projections used in Eq.~\ref{Eq:Chern}. Furthemore, we claim that the Chern number defined in Eq.~\eqref{Eq:Chern} is a topological invariant associated to the bundle traced by $\PM_G(\varphi,w)$ inside $B(\Hh_\Ss)$. Let us point out that, while $\PM_G(\varphi,w)$ cannot be expected to be compact operators, their derivations w.r.t. the parameters are expected to be compact.

Regarding the algebra generated by the models, we point out that states on $\Ss$ can be generated from any spectral gap (not just the ground state gap) that remains open during the thermodynamic limiting process.  By repeating the above construction, we can represent the corresponding gap projections as projections from $B(\Hh_\Ss)$. As such, the algebra that generates the gap projections of the models should be sought inside $B(\Hh_\Ss)$. In fact, it can be formally defined as the smallest $C^\ast$-subalgebra of $B(\Hh_\Ss)$ that contains all the gap projections of the models \eqref{Eq:GenHam4}, as represented inside $B(\Hh_\Ss)$. We believe this gives us a path towards a rigorous formalization of the topological invariants associated with the models \eqref{Eq:GenHam4} and a framework for applying the index theory developed in \cite{CareyAMS2014}.

Coming back to the more practical aspect of our work, we recall that there is an entire machinery \cite{ProdanJGP2019} for generating aperiodic lower dimensional systems that support topological phases from much higher dimensions. Using this machinery, for example, we can generate aperiodic 1-dimensional spin lattices which display the 4-dimensional integer quantum Hall physics in the $M=1$ sector. Using the same approach as in this paper, we can then generate a concrete class of many-body states that carry first and second Chern numbers. It will be interesting, for example, to test if the predictions from \cite{LuAOP2020} about the first Chern hold for these models. Also, since the first Chern numbers are weak invariants for these models, it will be also interesting to test their stability under strong disorder.

Lastly, we want to mention that we are presently working on generating topological many-body states from ones found in the $M=2$ and $M=3$ sectors in \cite{LiuArxiv2020}. This will supply a class of models where Theorem 2 from \cite{LuAOP2020} on the fractional Chern values can be tested.

	\acknowledgments{Both authors are supported by the NSF Grant No DMR-1823800. E.P. acknowledges additional financial support from the W.M. Keck Foundation. Both authors thank Lea. F. Santos for useful discussions and for supplying a generic computer code for spin systems.}

\end{document}